\documentclass[preprint]{aastex}
\usepackage{graphicx,natbib}

\newcommand{\ovi}{O~{\sc{vi}}}

\newcommand{\lya}{Ly$\alpha$}

\newcommand{\hi}{H~{\sc{i}}}

\slugcomment{Submitted to ApJ on December 16, 2013. Accepted on January 28, 2014.}

\shorttitle{Plasma physical parameters along CME-driven shocks: \\
I observations}
\shortauthors{Bemporad, Susino \& Lapenta}

\begin{document}

\title{Plasma Physical Parameters along Coronal Mass Ejection-driven Shocks: \\
I UV and White Light Observations}

\author{A. Bemporad, R. Susino}
\affil{Istituto Nazionale di Astrofisica (INAF), Osservatorio Astronomico di Torino, \\ Strada Osservatorio 20, 10025 Pino Torinese, Torino, Italy} 
\email{bemporad@oato.inaf.it}
\and
\author{G. Lapenta}
\affil{Center for Plasma Astrophysics (CPA), KU Leuven, \\ Celestijnenlaan 200B, 3001 Leuven, Belgium} 

\begin{abstract}
In this work UV and white light (WL) coronagraphic data are combined to derive the full set of plasma physical parameters along the front of a shock driven by a Coronal Mass Ejection. Pre-shock plasma density, shock compression ratio, speed and inclination angle are estimated from WL data, while pre-shock plasma temperature and outflow velocity are derived from UV data. The Rankine-Hugoniot (RH) equations for the general case of an oblique shock are then applied at three points along the front located between $2.2-2.6$~R$_\odot$  at the shock nose and at the two flanks. Stronger field deflection (by $\sim 46^\circ$), plasma compression (factor $\sim 2.7$) and heating (factor $\sim 12$) occur at the nose, while heating  at the flanks is more moderate (factor $1.5-3.0$). Starting from a pre-shock corona where protons and electrons have about the same temperature ($T_p \sim T_e \sim 1.5 \cdot 10^6$ K), temperature increases derived with RH equations could better represent the protons heating (by dissipation across the shock), while the temperature increase implied by adiabatic compression (factor $\sim 2$ at the nose, $\sim 1.2-1.5$ at the flanks) could be more representative of electrons heating: the transit of the shock causes a decoupling between electron and proton temperatures. Derived magnetic field vector rotations imply a draping of field lines around the expanding flux rope. The shock turns out to be super-critical (sub-critical) at the nose (at the flanks), where derived post-shock plasma parameters can be very well approximated with those derived by assuming a parallel (perpendicular) shock.
\end{abstract}


\section{Introduction}
The study of interplanetary shocks accelerated by Coronal Mass Ejections (CMEs) is very important to provide a better understanding of fundamental plasma physics processes involved, like the acceleration of energetic particles at the shock and wave-particle interactions replacing binary collisions in collisionless plasmas. After a long debate in the scientific community, it is now widely accepted that Solar Energetic Particles (SEPs - electrons and ions propagating at energies ranging from a few keV up to some GeV) are accelerated by two different sources (involving different acceleration physical mechanisms): solar flares (producing the so-called impulsive SEP events) and CME-driven shocks \citep[producing the so-called gradual events; see recent review by][]{reames2013}. Nevertheless, SEPs accelerated by interplanetary shocks are much more important regarding their space weather implications: particles accelerated in gradual events reach the highest energies and stronger fluxes, and due to the extension of interplanetary shock waves, these particles are also injected over a much broader region of the interplanetary space with respect to SEP accelerated in flares. Thus, the interaction of these particles with the Earth environment as they propagate along the interplanetary Parker spiral is much more common with respect to SEP associated with impulsive events, whose sources are clearly concentrated on the western half of the Sun magnetically connected with the Earth.

Nevertheless, the acceleration of SEPs by CME-driven shocks as well as their propagation in the interplanetary medium are still not well understood and one of the main open problems is the location in the corona of seed particles being accelerated. Over the last decades, much information on SEPs and associated interplanetary shock waves were derived from in situ data acquired from many different spacecrafts located at many different heliocentric distances, with the closest approach ever reached to the Sun around 0.29~AU, thanks to data acquired by Helios 1 and Helios 2 spacecrafts \citep[see e.g.][]{kallenrode 1993}. It has been pointed out that CME-driven shocks are likely most efficient in accelerating electrons in the heliocentric distance range between $1.5-4.0$~R$_\odot$ \citep[e.g.][]{gopalswamy2009a}, hence quite close to the Sun, in a region so far unexplored by in situ data. This could be due to a combination of the CME speed and the characteristic speeds of the medium crossed by the CME, leading to the production of strong shocks only closer to the Sun, while shocks became too weak or decayed by the time the CME reached the outer corona. This idea is in agreement with the observational evidence that type-II radio bursts (due to $\sim 10$~keV electron beams accelerated by the shocks and able to generate plasma waves at the local plasma frequency $f_{pe} \propto \sqrt{n_e}$) are excited only when CMEs are closer to the Sun. 

In fact, the theory of piston-driven shock waves induced in collisionless plasmas requires that the driver (i.e. the CME in this case) propagates in the medium (i.e. the solar corona) relatively faster than the local Alfv\'en or magnetosonic speeds. The Alfv\'en speed $v_A$ of a plasma with mass density $\rho$ permeated by a magnetic field $B$ is given by $v_A = B/\sqrt{\mu_0 \rho}$, while the magnetosonic speed $v_{ms}$ depends on the wave propagation angle $\theta$ with respect to the magnetic field and is given by $v_{ms} = \sqrt{v_A^2 + c_s^2}$ (with $c_s$ sound speed) only in the special case of perpendicular shock ($\theta = 90^\circ$). As CMEs propagate and expand from the lower corona to the interplanetary medium, they are expected to meet a plasma with a local minimum of $v_A$ (hence of $v_{ms}$) around $1.2 - 1.4$ R$_\odot$ and a local maximum around 3.5~R$_\odot$ \citep[see e.g.][]{mann2003}, and then $v_A$ progressively decays at higher distances (mainly because of the magnetic field radial decay) allowing the shock wave to survive very far from the Sun. The fundamental parameter controlling the strength of the shock is the Alfv\'enic shock Mach number $M_A$, given by the ratio of the upstream flow speed along the shock normal $v_{un}$ (in a reference frame at rest with the shock) to the upstream Alfv\'en speed $M_A = v_{un}/v_A$. It is well known that when the Mach number exceeds a certain (angular dependent) critical value $M_A^*$ the shock cannot be sustained by purely resistive dissipation like anomalous resistivity and viscosity alone. The excess energy is then rejected from the shock by reflecting part of the incoming plasma back up-stream. The up-stream plasma can thus cross multiple times the shock surface, being in turn accelerated up to SEP energies \citep[see e.g.][]{edminstonkennel1984}. Hence, the supercriticality of shocks is considered as a good indicator of their ability to accelerate particles and accurate methods for deducing shock strengths are indispensable, even if other authors pointed out that a determining parameter could also be the existence of seed supra-thermal particles located in the coronal regions crossed by the shock \citep[see e.g.][]{mason1999,lee2007}. 

Because, as mentioned, no in situ data are available close to the Sun, shock properties in the lower corona have, so far, only been explored with remote sensing data. Over the last decade, unique information on CME-driven shocks propagating in the corona have been derived from the analysis of White-Light (WL) coronagraphic images \citep[e.g.][]{vourlidas2003,rouillard2011}. These data have proven to be very useful to derive shock speeds, shock compression ratios $X = \rho_d / \rho_u$ \citep[i.e. the ratio between the up-stream $\rho_u$ and the down-stream $\rho_d$ densities; e.g.][]{ontiverosvourlidas2009}, and strengths of coronal magnetic fields crossed by the shocks \citep[e.g.][]{gopalswamyyashiro2011}, and allowed also statistical studies, for instance, on the correlation between the peak SEP intensities and associated CME speeds \citep[e.g.][]{kahler2001}. The study of decameter-hectometric to kilometric type-II radio bursts was also of fundamental importance, providing information about the shock compression ratios, shock speeds and strengths \citep[e.g.][]{mann2003,vrsnak2004,mancusoabbo2004}. Moreover, unique information (not available from the analysis of WL and radio data) on post-shock plasma heating and acceleration were provided by the analysis of UV spectra \citep[e.g.][]{raymond2000,mancuso2002}. Nevertheless, many information on shocks have been derived only when comparative analyses were performed by using remote sensing data acquired with very different wavelengths, like radio and UV \citep[e.g.][]{mancuso2002,mancusoavetta2008}, radio and WL \citep[e.g.][]{reiner2003}, and more recently UV and WL \citep[e.g.][]{bemporadmancuso2010}.

In this work we demonstrate how UV and WL data can be used to derive the full set of plasma physical parameters all along the front of a shock, including not only the strength of pre- and post-shock magnetic and velocity fields, but also the rotation induced by the transit of the shock in the magnetic and velocity field vectors. These information are of fundamental importance for our understanding of the physical processes occurring during the propagation of interplanetary shock waves and possibly to reveal the location of SEP acceleration in the corona. The CME-driven shock associated with the event studied here has already been analyzed by \citet[][]{bemporadmancuso2011} by using SOHO/LASCO WL data alone. In this work we extend the results previously obtained by also taking advantage of UV spectra acquired by the UV Coronagraph Spectrometer \citep[UVCS; see][]{kohl1995}. This paper is the first one in a sequence of two dealing with CME-driven shocks: the second one will focus on MHD simulations and comparison with observations. The paper is organized as follows: first (\S~2) we summarize for the reader's convenience the relevant techniques we developed and results we obtained in previous works and applied here, then we describe how data have been analyzed to derive the required up-stream plasma parameters from WL (\S~3.1) and UV (\S~3.2) data. In what follows, we explain how from these parameters the full set of down-stream plasma parameters has been derived with Rankine-Hugoniot (RH) equations (\S~4). Then, the obtained results are summarized and discussed (\S~5).

\section{Previous results of shocks with UV and WL data}

On 1999 June 11 a CME was observed by the LASCO~C2 and C3 coronagraphs \citep[][]{brueckner1995}; as reported by the standard ``CDAW'' CME catalog \citep[][]{gopalswamy2009b}, the event had a first order projected speed $v \sim 1570$~km~s$^{-1}$ (and hence can be classified as a fast event) with a central propagation angle by 35$^\circ$ (measured counter-clockwise from the solar north). The event was associated with a C8.8 flare and a type-II radio burst. At the day of the CME, the source Active Region (NOAA 8585) was located quite close to the solar limb (latitude of 38$^\circ$N, longitude of 64$^\circ$W), hence the CME expanded very close to the LASCO plane of the sky (hereafter POS). Given the large speed of the CME, the front was observed only in two LASCO~C2 frames acquired at 11:26 and 11:50~UT \citep[see Figure~\ref{fig01} here and Figure 1 in][, hereafter Paper~1]{bemporadmancuso2011}

In this work we extend the results already obtained by Paper~1 with LASCO~C2 WL data alone, where we demonstrate for the first time that the shock compression ratio $X = \rho_d / \rho_u$ between the up- and down-stream plasma densities can be measured from WL data all along the shock front. Results show that in both the analyzed LASCO~C2 frames the compression ratio maximizes at the center of the shock, but only a small region around the shock center is supercritical at earlier times, while higher up in the corona the whole shock becomes subcritical. In Paper~1 not only the shock compression ratio $X$ was derived all along the shock front, but also the up-stream coronal density $\rho_u$ along the front, from the last polarized brightness ($pB$) image acquired by LASCO~C2 before the CME (1999 June 10 at 21:00~UT) via classical Van-Der-Hulst inversion. Moreover, Paper~1 also derived all along the front the angle $\theta_{sh}$ between the normal to the shock front and the radial direction \citet[see also][for more details]{bemporadmancuso2013}. This angle can be considered as representative of the angle between the normal to the shock front and the up-stream magnetic field, in the assumption that the coronal field is nearly radial at these altitudes (see later). The same technique was also applied later on by \citet[][hereafter Paper~2]{bemporadmancuso2013} to another event in order to show that the absence of a super-critical region along the shock front also corresponds to a lack of type-II radio emission.

In this work, results already obtained for the same event in Paper~1 and Paper~2 are extended by taking advantage of the analysis previously developed by \citet[][hereafter Paper~3]{bemporadmancuso2010} for UV spectra acquired by SOHO/UVCS relative to another event. Hence, and we refer the interested reader to the cited papers for much more details. The main idea is to derive, after the analysis of WL data described above, the missing physical parameters of the up-stream plasma crossed by the shock from the analysis of UV data. In Paper~3 it was shown that, given all the up-stream plasma parameters except the magnetic field $B_u$ (i.e. given the density $n_u$, temperature $T_u$, velocity $v_u$, and the inclination angles between the shock normal and the up-stream magnetic and velocity field vectors, $\theta_{Bu}$ and $\theta_{vu}$, respectively), and given the compression ratio $X$, it is possible to apply the MHD RH equations for the general case of an oblique shock to derive not only all the down-stream physical parameters ($n_d$, $T_d$, $v_d$, $\theta_{Bd}$ and $\theta_{vd}$), but also the up-stream and down-stream magnetic fields $B_u$ and $B_u$. Nevertheless, a unique solution can be derived only in the regions of the ($\theta_{vu}$, $\theta_{Bu}$) plane where $\theta_{vu} \ll \theta_{Bu}$ or $\theta_{vu} \gg \theta_{Bu}$. The analysis we perform here is based on the same methods described in Paper~3, but applied to different points along the same shock front.

In Paper~3, due to the lack of significant emission in other spectral lines, standard techniques for the determination of plasma temperature from UV-EUV emission \citep[like the line ratio or differential emission measure techniques; see e.g.][and references therein]{parenti2000} have not been applied. Nevertheless, as it has been shown in Paper~3, pre-shock temperatures and densities can successfully be determined simply from the \ovi\ $\lambda$ 1031.91 \AA\ and \hi\ $\lambda$ 1215.67 \AA\ line intensities observed before the arrival of the shock front (i.e. the strongest lines available), once the effects due to Doppler dimming (see below) and to the integration along the line of sight (hereafter LOS) through the optically thin coronal plasma are both taken into account; for the same reasons this technique will also be applied here. The emission in these coronal spectral lines is mainly due to ion excitation by collisions with thermal coronal electrons (referred to as a collisional component) and absorption of chromospheric radiation (referred to as a radiative component), both followed by spontaneous emission \citep[see e.g. recent review by][]{bradshawraymon2013}. As it is well known, in the typical conditions of coronal plasma, collisional excitation for the \hi\ \lya\ line can be neglected \citep[][]{gabriel1971}, while both collisional and radiative excitation occur for the O~{\sc vi} $\lambda\lambda$ 1031.91 and 1037.61 \AA\ doublet lines. Collisional components mainly depend on the LOS distribution of electron density and temperature, while radiative components also depend on the plasma outflow velocity. When the plasma flow is not negligible, the scattering profile is Doppler-shifted with respect to the disk profile resulting in a less efficient scattering, hence in a reduction in intensity of the scattered radiation, an effect known as Doppler dimming \citep[][]{noci1987}. Thus the resonance O~{\sc vi} $\lambda\lambda$ 1031.91 and 1037.61 \AA\ line doublets are well suited to empirical determinations of the relative intensities of their collisionally and radiatively excited components, hence for the determination of the ion outflow velocity, as originally pointed out by \citet[][]{kohlwithbroe1982}.

Before starting the description of the new work, we add here to the previous publications a few short comments on the analysis method: first, even if compression ratios were derived from two-dimensional (2D) WL images (showing only the emission projected on the POS), the integration along the LOS crossing the three-dimensional (3D) structure of the expanding shock front was taken into account in the technique described in Paper~3. The assumption of a 3D paraboloidal-like shape for the shock front was previously validated by the comparison between WL images observed by SOHO -- STEREO coronagraphs and WL images simulated with forward modeling \citep[see e.g.][]{woodhoward2009,ontiverosvourlidas2009}. Second, we notice that values derived in Paper~1 for the projected thicknesses (0.06~R$_\odot$ at 11:26 UT and 0.09~R$_\odot$ at 11:50 UT) of the shock shell at different altitudes/times are in good agreement with values given by \citet[][Figure~5, bottom panel]{eselevich2012} and identified as representative of the proton mean free paths in the corona below heliocentric distances of 6~R$_\odot$. Third, these values for the thicknesses of the shock shell were used (after correction for the shock motion during the LASCO~C2 exposure time) to estimate reliable values of the actual shock depth $L$ along the LOS (0.28~R$_\odot$ at 11:26~UT and 0.61~R$_\odot$ at 11:50~UT), which is a critical parameter for the derivation of the compression ratios; this estimate was different to previous and successive works \citep[e.g.][]{ontiverosvourlidas2009,kim2012} where a constant shock depth $L = 1$ R$_\odot$ was simply assumed for all the events and all heliocentric distances.

\section{Observations and data analysis}

\subsection{Description and analysis of LASCO data}

In this work we first extend the analysis of WL LASCO data already performed in Paper~1 showing for the first time that WL images can also be employed in order to derive another important physical parameter all along the location of the shock front: the shock speed $v_{sh}$. Figure~\ref{fig02} shows the heliocentric distances $h_{sh}$ of points located at different latitudes identified along the shock fronts observed in the 11:26~UT (lower solid curve) and 11:50~UT (upper solid curve) LASCO~C2 frames (see also Figure~1 in Paper~1, right panels). These two curves can be easily employed to derive the latitudinal distribution of the average shock speed, by assuming isotropic self-similar expansion of the front in the range of common latitudes between the two curves. Recent results have shown that self-similar expansion is a realistic hypothesis for the evolution of CMEs in the intermediate and extended corona \citep[e.g.][]{mierla2011} and even for CME bubbles and CME voids in the lower corona \citep[e.g.][]{aschwanden2009,patsourakos_2010}. Nevertheless, the front could expand, in general, not symmetrically about the radial direction, and hence it is necessary to correct for the possible latitudinal propagation. To this end, we identified the center of the shock at 11:26~UT and 11:50~UT as the latitudinal location of the highest point along the front. Then, we determined the latitudinal shift required to have the center of the shock at the two times located at the same latitude. In particular, for this event the required latitudinal shift is quite small (4.7$^\circ$, see Figure~\ref{fig02}, left panel), but this angle could be much larger for other events.

After the correction for the latitudinal shock propagation, the average shock speed has been determined at each latitude simply as $v_{sh} = \Delta h_{sh} / \Delta t$. The resulting latitudinal distribution of shock speeds is shown in Figure~\ref{fig02} (right panel): the resulting speed is larger at the center of the shock, where it reaches a value around $v_{sh} \sim 1570 - 1580$~km~s$^{-1}$, and then it decreases towards the shock flanks, going down to $v_{sh} \sim 1340 - 1350$~km~s$^{-1}$ $\sim 15^\circ - 25^\circ$ away from the center. Notice that the maximum speed we derived is compatible with the CME speed of 1570~km~s$^{-1}$ as provided by the online LASCO CME catalog \citep[][]{gopalswamy2009b}. Hence, the analysis of white-light coronagraphic images can provide a lot of physical information on the shock; in summary, the up-stream coronal plasma density, the shock compression ratio $X$ (see Paper~1), the angle $\theta_{sh}$ between the normal to the shock front and the radial direction (see Paper~2), and the shock speed $v_{sh}$ (as shown here). In the next section we discuss how these parameters have been combined with other plasma parameters derived from the analysis of UV data.

\subsection{Description and analysis of UVCS data}  

As described above, from the analysis of WL data we already derived $n_u$ (from polarized measurements), $\theta_{Bu}$ and $\theta_{vu}$ (by simply assuming that the pre-shock wind and magnetic field are radial), the shock speed $v_{sh}$ and the compression ratio $X$. Hence, the only missing physical parameters that will be derived here from the analysis of UV data are the up-stream temperature $T_u$ and the coronal outflow velocity $v_{out}$. The latter quantity is needed in order to derive the quantities $v_u$ and $\theta_{vu}$ written in a reference system at rest with the shock front. To this end, the so-called UVCS synoptic observations are really helpful. Under this program, a fixed sequence of exposures were acquired every day over $\sim 12$~hours for 8 different polar angles (separated by angular steps of $\pi /4$) and covering heliocentric distances between 1.5 and 3.0~R$_\odot$, in order to map the UV emission in the whole intermediate corona. These data have already been successfully employed, for instance, to perform 3D tomographic reconstructions of the UV solar corona \citep[e.g.][]{panasyuk1999}, to study the period of UV coronal rotation \citep[see e.g.][]{giordanomancuso2008}, and to provide Carrington maps of coronal plasma physical parameters \citep[e.g.][]{ko2008,strachan2012}.

\begin{figure}[th]
\epsscale{.90}
\plotone{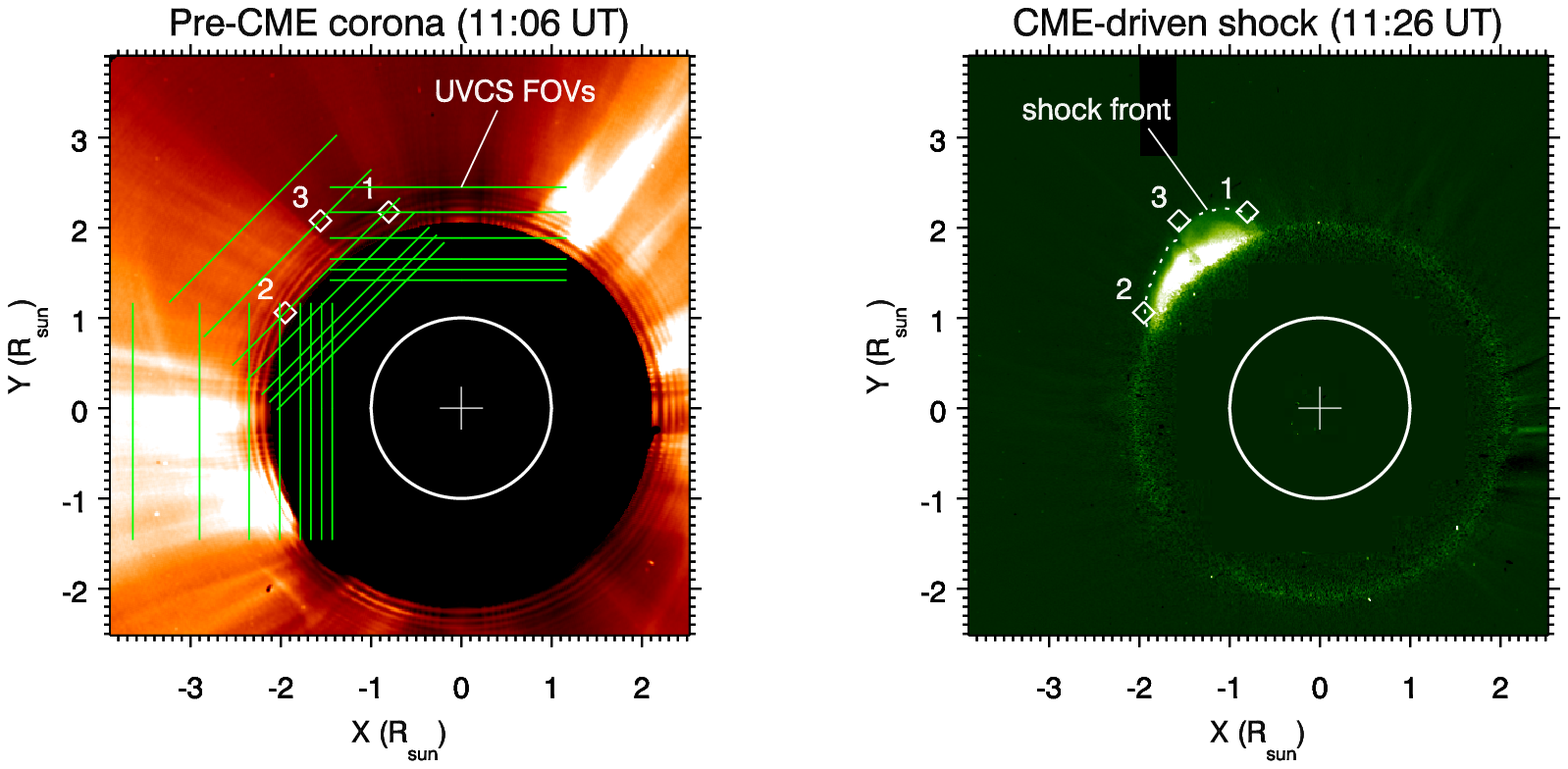} 
\newline
\newline
\newline
\caption{
Left: LASCO~C2 raw image showing the pre-CME corona at 11:06 UT. The image has been normalized to the average white light radial profile extracted in the same frame above the polar coronal holes, in order to show the location of brighter structures. The panel also shows the location of the UVCS slit field of views for the synoptic observations acquired before the CME (yellow solid lines, see text), as well as the location (diamonds) of the three points in the corona where the shock has been analyzed here (points 1, 2 and 3, see text). Right: LASCO~C2 11:26 UT base difference image (frame at 11:06 UT subtracted) showing the location of the CME-driven shock front with respect to the three points being analyzed here (same as in the left panel).
} \label{fig01}
\end{figure}

In this work, synoptic observations acquired before the CME have been analyzed in order to provide the required missing pre-shock plasma temperatures and outflow velocities: these are assumed to be representative of the up-stream parameters met by the CME-driven shock, by assuming that no significant coronal evolution occurred in a few hours before the CME. The UVCS field of views for synoptic data acquired on 1999 June 11 in the north-east quadrant are summarized in Figure~\ref{fig01}; all these data were acquired just before the CME. In particular, the sequence with the slit center above the north pole was acquired between 06:17:46 and 07:50:01~UT (26 exposures acquired with the slit center at 1.418, 1.536, 1.654, 1.886, 2.171, and 2.450~R$_\odot$); the sequence with the slit center at a latitude of 45$^\circ$ was acquired between 07:55:27 and 09:28:35~UT (26 exposures acquired with the slit center at 1.432, 1.550, 1.669, 1.901, 2.129, 2.576, and 3.117~R$_\odot$). Finally, the sequence with the slit center at the equator was acquired between 09:34:07 and 11:49:13~UT (38 exposures acquired with the slit center at 1.430, 1.548, 1.666, 1.783, 2.013, 2.352, 2.901, and 3.640~R$_\odot$). 

A summary of the projected locations of the UVCS slit fields of view for synoptic observations described above is reported also in Figure~\ref{fig03} in a polar plot. All these data have been acquired in the so-called ``O~{\sc vi}-channel'' with exposure times between $180 - 200$~s (depending on the slit projected altitude), slit width of 100~$\mu$m, spectral resolution of 0.2979~\AA/pixel and spatial resolution of 21''/pixel. As usually done for UVCS synoptic observations, two spectral ranges were included: a first range between $1023.43- 1043.68$~\AA\ and a second range between $975.71 - 992.39$~\AA\ (corresponding to the range between $1207.56 - 1222.93$~\AA\ on the redundant channel). The strongest spectral lines included in these ranges are: H~{\sc i} Lyman-$\alpha$ $\lambda 1215.67$ \AA\ and Lyman-$\beta$ $\lambda 1025.72$ \AA\ lines, the O~{\sc vi} $\lambda\lambda$ 1031.91 and 1037.61 \AA\ doublet lines, and the Si~{\sc xii} $\lambda 520.66$ \AA\ and Mg~{\sc x} $\lambda 609.86$ \AA\ second order lines.

Figure~\ref{fig03} shows that, once the location of the CME-driven shock front (identified from LASCO~C2 frame acquired at 11:26~UT) is over plotted on the location of synoptic observations, a few intersections occur. In particular, if one eliminates the crossings occurring too close to the slit edges (where line intensities could be subject to larger uncertainties), three coronal up-stream regions can be investigated along the shock front (indicated in Figure~\ref{fig03} by triangles). These three regions are very suitably located at the right (i.e., northward) and left (i.e., equatorward) flanks of the shock (points labeled as ``1'' and ``2'', respectively, in Figure~\ref{fig03}) and the center of the shock (point ``3'' in Figure~\ref{fig03}). Hereafter, the analysis will focus only over these three points located at three different latitudes ($\theta_1 = 69.7^\circ$, $\theta_2 = 28.5^\circ$, $\theta_3 = 53.1^\circ$) and heliocentric distances ($h_1 = 2.32$ R$_\odot$, $h_2 = 2.22$ R$_\odot$, and $h_3 = 2.60$ R$_\odot$) along the front; in what follows, we will then use the above numbering to refer to these different points. The location of the three points in the corona is also shown in Figure~\ref{fig01} (left panel) projected over the last frame acquired by LASCO~C2 before the arrival of the CME.

The UVCS synoptic data being analyzed here have been acquired with the slit centered at heliocentric distances and latitudes ($h$, $\theta$) of (2.171 R$_\odot$, 90$^\circ$N), (2.129 R$_\odot$, 45$^\circ$N), and (2.576 R$_\odot$, 45$^\circ$N), for the study at coronal points 1, 2 (shock flanks) and 3 (shock center), respectively. For each one of these locations a total of 6, 4, and 5 exposures have been acquired, all with exposure times of 180~s. Figure~\ref{fig04} shows the average spectra (in two different spectral intervals) observed at each point summing over all the available exposures. Moreover, in order to increase the signal-to-noise ratio, a further average over 8~bins in the spatial direction has been performed. The spectra shows clearly that only 3 spectral lines are detected with a good statistic: the O~{\sc vi} $\lambda\lambda$ 1031.91 and 1037.61 \AA\ doublet lines (top panels) and the H~{\sc i} Lyman-$\alpha$ $\lambda 1215.67$ \AA\ (bottom panels). These are the strongest spectral lines emitted in the UV range by coronal plasmas, hence the lines with the best counting statistics.

As previously described, we then determined the pre-shock coronal outflow velocities $v_{out}$ from the detected ratios between the O~{\sc vi} $\lambda\lambda$ 1031.91 and 1037.61 \AA\ doublet lines, with a technique well-established in the analysis of UVCS data \citep[see e.g. reviews by][]{antonucci2006,kohl2006}. Outflow velocities of the O$^{5+}$ ions turn out to be $v_{out1} = 41.7$~km~s$^{-1}$ and $v_{out2} = 61.7$~km~s$^{-1}$ for points 1 and 2 (shock flanks), and $v_{out3} = 70.4$~km~s$^{-1}$ for point 3 (shock nose), respectively. The pre-shock coronal outflow velocity values we determined are consistent with velocities expected for O$^{5+}$ ions at the considered range of heliocentric distances ($\sim 2.3 - 2.6$~R$_\odot$) in coronal streamers \citep[see e.g.][]{nocigavryuseva2007}, also taking into account that on 1999 June 11 WL coronal images acquired by the LASCO~C2 coronagraph before the CME (Figure~\ref{fig01}, left panel) show that points 1 and 3 are located at lower altitudes in brighter (hence denser) coronal regions, where lower outflow velocities are expected, while point 3 is located at slightly higher altitude in a darker coronal region.

The above velocities have then been used to estimate the Doppler dimming factors for both the \ovi\ $\lambda$ 1031.91 \AA\ and \hi\ $\lambda$ 1215.67 \AA\ radiative components, by assuming that O$^{5+}$ and p$^{+}$ ions have the same outflow velocities. This assumption seems reliable for observations relative to heliocentric distances not larger than 3 R$_\odot$, as those considered here \citep[see discussion by][about their Figure 41, and references therein]{kohl2006}. In the computation we also assumed typical ion kinetic temperatures given by \citet[][]{abbo2010} at different heliocentric distances (region 1, internal streamer) by neglecting as a first approximation the ion temperature anisotropy. These quantities affect the widths of atomic absorption profiles, hence the resulting value of Doppler dimming factors. Then, by applying the same analysis performed in Paper~3, distribution functions for the unknown electron temperature, electron density, and plasma velocity ($T_e(z)$, $n_e(z)$, and $v_{out}(z)$) along the LOS coordinate $z$ have been assumed from \citet{cranmer1999}. The $v_{out}(z)$ profile has been normalized to the POS $v_{out}$ values given above, while $T_e(z)$ and $n_e(z)$ profiles have been multiplied by two independent constant factors $\left(K_T, K_n \right)$, respectively. By computing the expected \ovi\ $\lambda$ 1031.91 \AA\ and \hi\ $\lambda$ 1215.67 \AA\ radiative and collisional intensities integrated along the LOS for a range of $\left(K_T, K_n \right)$ pair values and comparing them with the observed line intensities, values of $\left(K_T, K_n \right)$ best reproducing the UV observations have been determined. In particular, derived values are $\left(K_T, K_n \right)_1 = \left(1.50, 2.70 \right)$, $\left(K_T, K_n \right)_2 = \left(0.86, 5.21 \right)$ and $\left(K_T, K_n \right)_3 = \left(1.50, 3.22 \right)$ for points 1, 2 and 3 respectively. This provides us with the LOS temperature and density profiles (given by $K_T\,T_e(r)$ and $K_n\,n_e(r)$) at each point in the corona.

\begin{table}[t]
\begin{center} 
\begin{tabular}{|l|lllll|llll|}
\hline
 & \multicolumn{5}{|c|}{Observed quantities} & \multicolumn{4}{|c|}{Derived quantities}\\
\hline
P & $h$ & $\theta$ & $I_{1216}$ (ph s$^{-1}$ & $I_{1032}$ (ph s$^{-1}$ & $r_{OVI}$ & $v_{out}$ (km & $T_{e}$ & $n_{e}$ (UV) & $n_{e}$ (WL) \\
 & (R$_\odot$) &  ($^\circ$) & cm$^{-2}$sr$^{-1}$)& cm$^{-2}$sr$^{-1}$) & & s$^{-1}$) & (MK) & (cm$^{-3}$) & (cm$^{-3}$) \\
\hline 
1 & 2.32 & 69.7 & $1.08\cdot 10^{10}$ & $1.52\cdot 10^{8}$ & 3.54 & 41.7 & 1.68 & $4.22\cdot 10^{5}$  & $3.14\cdot 10^{5}$ \\
2 & 2.22 & 28.5 & $4.45\cdot 10^{10}$ & $1.16\cdot 10^{9}$ & 3.03 & 66.7 & 1.02 & $1.05\cdot 10^{6}$  & $7.59\cdot 10^{5}$ \\
3 & 2.60 & 53.1 & $6.75\cdot 10^{9}$ & $7.76\cdot 10^{7}$ & 2.87 & 70.4 & 1.49 & $2.95\cdot 10^{5}$  & $2.55\cdot 10^{5}$ \\
\hline
\end{tabular}
\end{center}
\caption{
Observed pre-shock coronal quantities (left) and derived plasma physical parameters (right) for the three coronal points shown in Figure~\ref{fig01} (left panel). Coronal quantities listed on the left part are the intensities of \ovi\ $\lambda$ 1031.91 \AA\ ($I_{1032}$), \hi\ $\lambda$ 1215.67 \AA\ ($I_{1216}$) spectral lines, and the ratio ($r_{OVI}$) between the \ovi\ $\lambda\lambda$ 1031.91 and 1037.61 \AA\ line intensities detected at different heliocentric distances ($h$) and latitudes ($\theta$) in the corona corresponding to the 3 points ($P$) considered in this work (Figure~\ref{fig01}). Derived quantities listed on the right part are the outflow speed of O$^{5+}$ ions ($v_{out}$), the electron temperature ($T_e$) and electron densities derived from WL ($n_e$ WL) and UV ($n_e$ UV) data.
} \label{tab01}
\end{table}

Resulting POS electron temperatures and densities derived with this technique are summarized in Table~\ref{tab01}. Results in this Table show that point 1 has an intermediate electron density $n_e$, consistent with its lower-altitude, but higher-latitude location; point 2 has higher $n_e$, in agreement with the brighter coronal appearance observed in WL at lower latitude and altitude; point 3 has the lowest $n_e$, in agreement with higher altitude and its location in a darker WL coronal region (see Figure~\ref{fig01}, left panel). Densities derived by UVCS data are in quite good agreement (within less than 30\%) with those derived independently from WL polarization measurements and are intermediate between typical densities measured in coronal streamers \citep[e.g.][]{gibson1999} and in coronal holes \citep[e.g.][]{cranmer1999}. This is also in agreement with Figure~\ref{fig01} that shows above the NE quadrant multiple structures fainter in WL than surrounding coronal streamers, and brighter than surrounding coronal holes. Derived electron temperatures are also compatible with typical streamer/coronal hole values at the corresponding altitudes and latitudes. 

Notice that temperatures derived through the above technique (and more in general from UV line intensities) are representative only of electron temperatures $T_e$ that could, in general, be different from proton temperatures $T_p$. Thanks to the tight coupling between neutral and ionized hydrogen by charge transfer, the latter plasma parameter can be derived from the FWHM of the \hi\ \lya\ line profiles (Figure~\ref{fig04}) with standard Gaussian fitting, after correction for the UVCS instrumental line broadening \citep[performed with the empirical formula given by][]{kohl1999}. Resulting proton temperatures $T_p$ for coronal regions considered here turn out to be between $1.0-1.5 \cdot 10^6$ K, quite similar to the electron temperatures $T_e$ derived above. Hence, at these altitudes the pre-shock coronal plasma seems to be in first approximation collisionally coupled ($T_p \simeq T_e$). Usually UVCS detected $T_p \simeq T_e$ in coronal streamers \citep[e.g.][]{li1998} and $T_p > T_e$ in coronal holes \citep[e.g.][]{kohl1997}, but this result is not universally accepted and other authors reported larger $T_p$ at streamer cores \citep[e.g.][]{zangrilli1999,akinari2007}. In fact, as it was demonstrated by \citet[][]{labrosse2006}, the reliability of proton temperatures derived by \hi\ \lya\ line profiles is limited, and derived $T_p$ can be significantly underestimated. For these reasons, in what follows we proceed by considering that our observational result that $T_e \simeq T_p$ in the pre-shock plasma is correct. In any case, we verified for instance that an increase by a factor 2 in the assumed pre-shock $T_p$ values would simply lead to an increase by less than a factor 2 in the resulting post-shock $T_p$. On the other hand, possible more significant differences between $T_p$ and $T_e$ in the post-shock plasma are discussed in the Conclusions.

\section{Determination of post-shock plasma parameters}

Once the plasma temperatures and outflow velocities are known (from UV data), together with densities (from UV and WL data), shock velocities, and shock inclination angles (from WL data), the MHD RH equations for the general case of oblique shock can be applied. These equations are valid under some well-known hypotheses: 1) the shock is stationary, 2) the curvature of shock surface can be neglected, 3) the flow and magnetic field directions in front and behind the shock front lie in the same plane, and 4) the energy transported by waves is negligible. As already demonstrated in Paper~3, these equations can be applied to derive the plasma down-stream parameters even if the up-stream magnetic field is unknown, when a measurement for the density compression ratio $X = \rho_d / \rho_u$ is provided in turn. The RH equations express simple conservation across the shock surface $S$ of mass, $x-$momentum, $y-$momentum, and energy flux, plus two more equations expressing the consequences of Poisson's equation ($ \nabla\cdot \mathbf{B} = 0$) and of Faraday's law ($\nabla\times \mathbf{E} = -\partial \mathbf{B}/\partial t$) across the shock surface. These conditions are usually written in a reference frame comoving with the shock surface $S$, with the $x$ and $y$ coordinated axes, respectively, perpendicular and parallel to $S$.

As previously done in Paper~3, conversion to this reference frame has been performed by assuming that, at the heliocentric distances of the coronal region considered here (i.e. between $\sim 2.2-2.6$~R$_\odot$), the magnetic field and outflow velocity are both radial. This is not a strong assumption, because it is well known from many total solar eclipse observations that coronal structures (outlining the magnetic field orientation) appear to be nearly radial above heliocentric distances of $\sim 2$ R$_\odot$. According to this observational fact, potential field reconstructions usually assume that above a spherical source surface (typically with a radius around 2.5~R$_\odot$) all field lines are radial \citep[see e.g.][and references therein]{saez2007}. This is also qualitatively confirmed in our case by the almost radial alignment of coronal structures visible at least in the north-east quadrant of the solar corona before the CME (Figure~\ref{fig01}, left). With this assumption, given the pre-shock coronal outflow velocities $v_{out}$ (from UV data), the shock velocities $v_{sh}$ (from WL data) and the inclination angles $\theta_{sh}$ between the normal to the shock front and the radial direction, it is simple to derive the vector components of the up-stream velocity $v_u$ in the reference frame at rest with $S$. In particular, with measured angles $\theta_{sh1} = - 54.5^\circ$, $\theta_{sh2} = + 32.6^\circ$, and $\theta_{sh3} = + 5.17^\circ$ (with positive values running counter-clockwise on the POS) for points~1,~2 (shock flanks) and~3 (shock nose), the up-stream velocities turn out to be $v_{u1} = 1328.3$~km~s$^{-1}$, $v_{u2} = 1281.2$~km~s$^{-1}$, and $v_{u3} = 1455.0$~km~s$^{-1}$. 

Unfortunately, as also pointed out in Paper~3, the number of possible solutions for $X$ provided by the RH equations needs to be discussed on the ($\theta_{vu},\theta_{Bu}$) plane, because in specific regions of that plane the solution for $X$ is not unique. In general, three solutions are possible (corresponding to fast, slow, and intermediate shock) in a region of the ($\theta_{vu},\theta_{Bu}$) plane closer to the straight line $\theta_{vu} = \theta_{Bu}$, while a unique solution is provided when $\theta_{vu} \ll \theta_{Bu}$ and $\theta_{vu} \gg \theta_{Bu}$ (see Paper~3, Figure 10, top left panel). Once the up-stream velocity vectors are converted in the reference frame at rest with $S$, it turns out that their inclination angles with respect to the normal to the shock are $\theta_{vu1} = - 1.46^\circ$, $\theta_{vu2} = 1.61^\circ$, and $\theta_{vu3} = 0.250^\circ$. On the other hand, the up-stream magnetic field inclination angles (as derived from WL images, by assuming that the magnetic field is radial) are $\theta_{Bu1} = - 54.5^\circ$, $\theta_{Bu2} = 32.6^\circ$, and $\theta_{Bu3} = 5.17^\circ$. Hence, for all the three points is $|\theta_{Bu}| \gg |\theta_{vu}|$ and unique solutions for $X$ exists, as we verified.

\begin{table}[t]
\begin{center} 
\begin{tabular}{|l|llllll|llll|}
\hline
P & $T$  & $n$ & $B$ & $v$ & $\theta_B$ & $\theta_v$ & $X$ & $v_{A}$ & $v_{s}$ & $v_{out}$ \\
 & (MK) & (cm$^{-3}$) & (G) & (km~s$^{-1}$) & ($^\circ$) & ($^\circ$) &  & (km~s$^{-1}$) & (km~s$^{-1}$) & (km~s$^{-1}$) \\
\hline 
1 ($u$) & 1.68 & $5.22\cdot 10^{5}$ & 0.241 & 1328.3 & -54.5 & -1.46  & 1.77 & 809.3 & 215.1 & 41.7 \\
1 ($d$) & 5.15 & $7.49\cdot 10^{5}$ & 0.415 & 794.0 & 70.3 & 19.6 & - & 1048.2 & 376.7 & 638.3 \\
\hline
2 ($u$) & 1.02 & $1.05\cdot 10^{5}$ & 0.511 & 1281.2 & 32.6 & 1.61  & 1.24 & 1089.6 & 167.6 & 66.7 \\
2 ($d$) & 1.51 & $1.30\cdot 10^{6}$ & 0.630 & 1078.6 & -46.8 & 17.0 & - & 1203.6 & 203.7 & 433.2 \\
\hline
3 ($u$) & 1.49 & $2.95\cdot 10^{5}$ & 0.205 & 1455.0 & 5.17 & 0.250  & 2.73 & 824.2 & 202.6 & 70.4 \\
3 ($d$) & 18.4 & $8.06\cdot 10^{5}$ & 0.330 & 766.8 & -51.7 & 45.9 & - & 802.7 & 711.1 & 967.5 \\
\hline
\end{tabular}
\end{center}
\caption{
Summary of plasma physical parameters (velocities in the reference frame at rest with the shock surface) for the three coronal points shown in Figure~\ref{fig01} up- ($u$) and down-stream ($d$), together with other plasma parameters like the shock compression ratio $X$, and the Alfv\'en ($v_A$), sound ($v_s$) and outflow ($v_{out}$) velocities in a reference frame at rest with the Sun. 
} \label{tab02}
\end{table}

Resulting up- ($u$) and down-stream ($d$) plasma physical parameters are summarized in Table~\ref{tab02} for the two points at the shock flanks (points~1 and~2) and for the shock nose (point~3). Table~\ref{tab02} reports the up- and down-stream velocities and vector angles in the reference frame at rest with the shock surface, together with the Alfv\'en ($v_A$), sound ($v_s$), and outflow ($v_{out}$) velocities in a reference frame at rest with the Sun. First of all, the Table shows that the up-stream velocity is larger everywhere than the magnetosonic velocity $v_{ms} = \sqrt{v_A^2+v_s^2}$, the shock is fast at both flanks and at the nose, hence a magnetic field compression occurs everywhere. Nevertheless, the stronger rotation of the coronal field occurs at the nose, where the shock is also stronger and larger plasma compression and heating are inferred. At the shock nose, the temperature increase with respect to the pre-CME corona is on the order of a factor $\sim 12$, hence larger than the temperature increase inferred in Paper~3 for a shock at 4.1 R$_\odot$; more moderate, but still significant (factor $1.5-3.0$) plasma heating occurs also at the shock flanks. The magnetic field is rotated clockwise at the shock nose and at the southward flank, while a counter-clockwise rotation occurs at the northward flank. The plasma has $\beta \ll 1$ up- and down-stream both at the shock flanks and at the nose, being significantly closer to unity ($\beta \simeq 0.47$) only in down-stream plasma at the nose, because of significant heating occurring there. The Table also shows that (in a reference frame at rest with the Sun) the plasma in the pre-shock corona is not only sub-alfv\'enic ($v_{out} < v_A$), but also sub-sonic ($v_{out} < v_s$). After the shock transit at the flanks (points~1 and~2) the plasma is accelerated up to super-sonic ($v_{out} > v_s$), but still sub-alfv\'enic ($v_{out} < v_A$) speed, and only at the shock nose (point~3) the plasma is accelerated up to super-alfv\'enic (but still sub-magnetosonic) speed. In the reference frame at rest with the Sun also strong deflections of the velocity vectors also occur. Starting from the velocity vector angles measured pre-shock with respect to the shock normal (equal to the angles $\theta_B$ given in Table~\ref{tab02} for the up-stream plasma), it turns out that after the shock transit the velocity vector rotates by $-29.8^\circ$, $-14.1^\circ$, and $-29.6^\circ$, respectively for points~1,~2 (shock flanks), and~3 (shock nose). Hence, a clockwise rotation of the velocity vector occurs everywhere along the front, at variance with the result for the magnetic field. It is worth noting that in this work it was assumed that field line deflections described above occur in the POS: this assumption will be discussed and justified in the last section.

It is also interesting to compare (Table~\ref{tab03}) the derived down-stream plasma parameters with those expected (given the same up-stream parameters) from the RH equations written in the simpler cases of parallel (``$\parallel$'') and perpendicular (``$\perp$'') shocks. Comparisons between shock compression ratios $X$ measured from WL data and those expected for parallel ($X_\parallel$) and perpendicular ($X_\perp$) shocks clearly show that at the shock flanks $X \simeq X_\perp$, while at the shock nose $X$ is much closer to the expected $X_\parallel$ value. The same similarity is present between the post-shock temperatures ($T_d$) derived at flanks and those expected for a perpendicular shock ($T_{d\perp}$), while temperatures at the nose are much better approximated by those expected for a parallel shock ($T_{d\parallel}$). Also, shock Mach numbers ($M_A$) measured at the flanks are very well approximated (for low plasma $\beta$ values) by those expected for a perpendicular shock ($M_{A\perp} = \sqrt{X(X+5)/[2(4-X)]}$), while the Mach number at the nose is much better approximated with the value expected for a parallel shock ($M_{A\parallel} = \sqrt{X}$). For future reference, this Table also reports the plasma temperatures expected by a simple adiabatic compression $T_{d\gamma} = T_u X^{(\gamma - 1)}$ (where $\gamma = 5/3$ is the adiabatic index, valid for fast hence adiabatic transitions across the shock).

Interestingly, Table~\ref{tab03} also confirms that measured $M_A$ values are quite well approximated by the empirical formula introduced in Paper~1 for the estimate of the Mach numbers $M_{A\angle}$ in the general case of an oblique shock
\begin{equation}
M_{A\angle} = \sqrt{\left(M_{A\perp} \sin \theta_{Bn} \right)^2 + \left(M_{A\parallel} \cos \theta_{Bn} \right)^2}
\end{equation} 
where $\theta_{Bn}$ is the angle between the up-stream magnetic field and the normal to the shock surface derived from WL data (see above). Hence, this formula can be used to provide approximated estimates for $M_A$ all along the shock fronts given the quantities $X$ and $\theta_{Bn}$ derived from WL images alone. Values for Mach numbers are also compared in Table~\ref{tab03} with the corresponding values for the first critical Mach number $M_A^*$, which represents the threshold $M_A$ value below which the down-stream velocity is sub-sonic \citep[][]{edminstonkennel1984}. Given the similarity between $M_A$ values measured in this work and approximate $M_{A\angle}$ values already estimated in Paper~1 with the above empirical formula, we also confirm here with a different technique the result already found: the shock is super-critical ($M_A \sim M_{A\angle} > M_A^*$) at the nose and sub-critical ($M_A \sim M_{A\angle} < M_A^*$) at the flanks. The criticality (subcriticality) of the shock at the nose (at the flanks) is also demonstrated by the velocities given in Table~\ref{tab02}, in the reference frame at rest with the shock: at the shock flanks $M_A < M_A^*$, and the normal component of the down-stream flow $v_{nd} = v_d \cos \theta_v > v_s$, hence the flow is still supersonic (and clearly sub-magnetosonic). On the other hand, at the shock nose $M_A > M_A^*$ and $v_{nd} < v_s$ because only above the critical Mach number the down-stream flow velocity should fall below the down-stream sound speed. Different Mach numbers for the perpendicular, parallel, and oblique shock cases, together with critical Mach numbers, are also plotted all along the shock front in Figure~\ref{fig05}. Notice that the condition $v_{nd} < v_s$ is verified at the nose not only because of the strong flow velocity decrease and deflection (in the shock reference frame) across the shock, but also because of the strong plasma heating leading to a significant increase down-stream of the sound speed $c_s = \sqrt{\gamma p/\rho} = \sqrt{\gamma k_B T /m_p}$.
 
\begin{table}[t]
\begin{center} 
\begin{tabular}{|l|lll|}
\hline
Point & 1 (N shock flank)  & 2 (S shock flank) & 3 (shock nose) \\
\hline
$T_u$  & $1.68 \cdot 10^6$ & $1.02 \cdot 10^6$ & $1.49 \cdot 10^6$ \\
$n_u$  & $4.22 \cdot 10^5$ & $1.05 \cdot 10^6$ & $2.95 \cdot 10^5$ \\
\hline
$X$ & 1.77 & 1.24 & 2.73 \\
$X_\perp$ & 1.72 & 1.21 & 1.86 \\
$X_\parallel$ & 3.71 & 3.80 & 3.78 \\
\hline
$T_{d}$ & $5.15 \cdot 10^6$ & $1.51 \cdot 10^6$ & $1.84 \cdot 10^7$ \\
$T_{d\perp}$ & $4.41 \cdot 10^6$ & $1.27 \cdot 10^6$ & $5.63 \cdot 10^6$ \\
$T_{d\parallel}$ & $2.15 \cdot 10^7$ & $1.95 \cdot 10^7$ & $2.53 \cdot 10^7$ \\
$T_{d\gamma}$ & $2.46 \cdot 10^6$ & $1.18 \cdot 10^6$ & $2.91 \cdot 10^6$ \\
\hline
$M_A$ & 1.64 & 1.17 & 1.76 \\
$M_{A\perp}$ & 1.65 & 1.19 & 2.88 \\
$M_{A\parallel}$ & 1.33 & 1.11 & 1.65 \\
$M_{A\angle}$ & 1.55 & 1.13 & 1.67 \\
\hline
\end{tabular}
\end{center}
\caption{
Comparison between the parameters derived for the general case of an oblique shock and those expected for the special cases of parallel (``$\parallel$'') and perpendicular (``$\perp$'') shocks (see text).
} \label{tab03}
\end{table}

\section{Discussion and Conclusions}

\begin{figure}[th]
\epsscale{.90}
\plotone{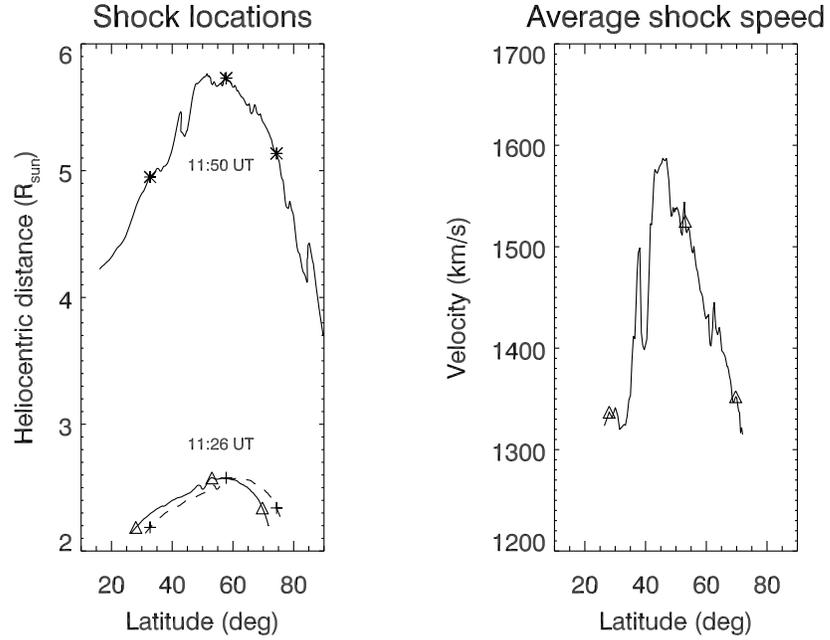}
\vspace{-9cm}
\caption{
Left: heliocentric distances of points located at different latitudes along the shock fronts identified at 11:26 UT (bottom solid line) and 11:50 UT (top solid line). The dashed line shows the same front shifted northward for the shock velocity estimate, while stars and triangles give the latitudinal location of the three points along the front where the full plasma parameters have been derived (see text). Right: latitudinal distribution of the average shock speed in the interval 11:26 -- 11:50 UT as derived from the curves shown in the left panel. Again, triangles show the location of the three points along the front where the full plasma parameters have been derived (see text).
} \label{fig02}
\end{figure}

\begin{figure}[th]
\epsscale{1.0}
\plotone{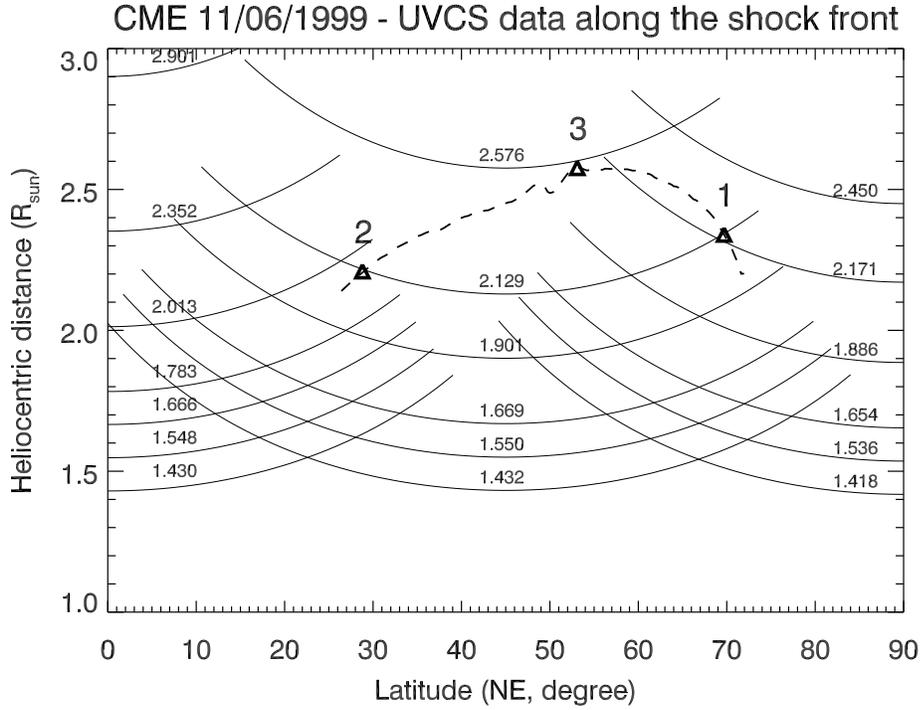}
\vspace{-11cm}
\caption{
Polar plot showing the locations of the UVCS fields of view during the synoptic observations acquired on 1999 June 11 before the transit of the CME-driven shock (numbers in the plot give the corresponding heliocentric distance of the UVCS slit center, in solar radii). The dashed line shows the position of the shock front as determined from the analysis of the LASCO~C2 frame acquired at 11:26 UT, while the triangles represent the location of the three points along the front where the full plasma parameters have been derived (see text).
} \label{fig03}
\end{figure}

\begin{figure}[th]
\epsscale{1.0}
\plotone{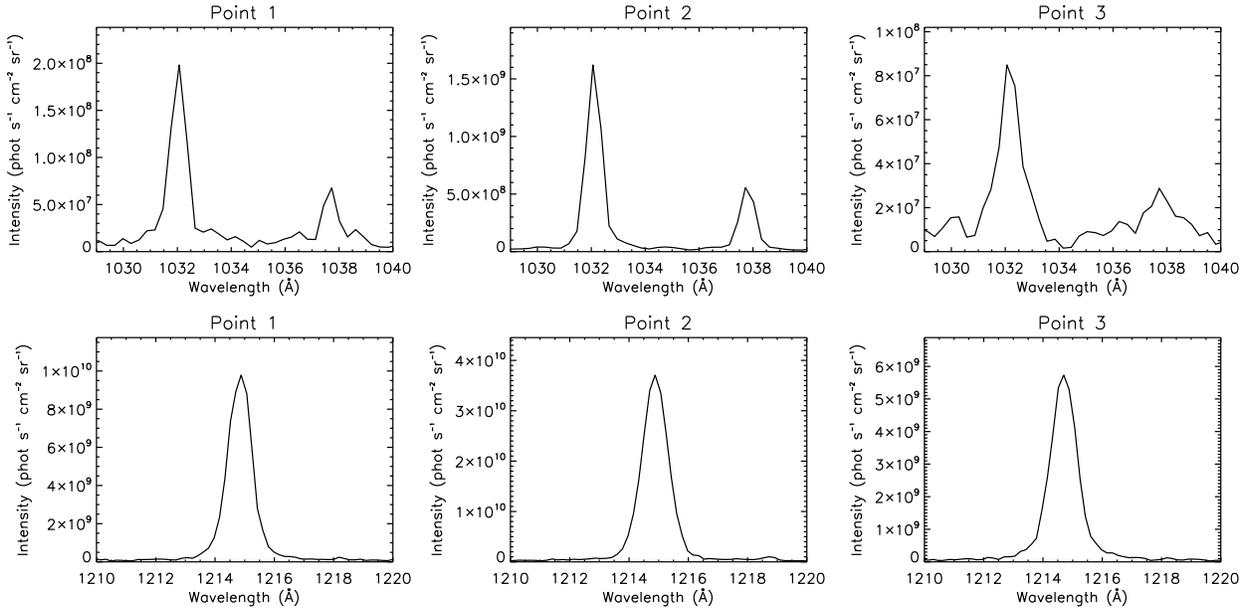}
\caption{
Top: O~{\sc vi} $\lambda \lambda$ 1031.91 and 1037.61 \AA\ average line profiles observed in pre-shock corona at the regions being crossed later on by the shock flanks (points $1 - 2$, left and middle panels) and the shock nose (point 3, right panel). Bottom: H~{\sc i} \lya $\lambda$ 1215.67 average line profiles (same order as in the top row).
} \label{fig04}
\end{figure}

\begin{figure}[th]
\epsscale{0.6}
\plotone{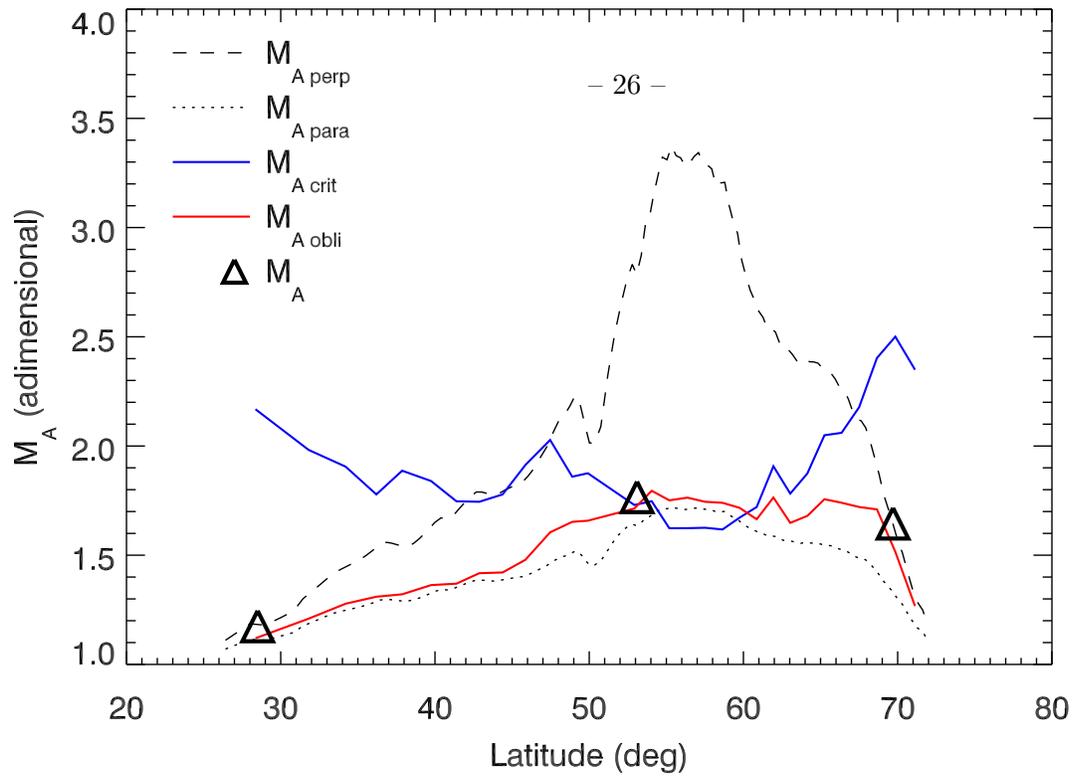}
\newline
\newline
\newline
\caption{
Values measured in Paper~1 for the Alfv\'enic Mach numbers in the case of perpendicular (dashed black line) and parallel (dotted black line) shocks, together with values for the critical Mach number (solid blue line). The Figure shows the very good agreement between values computed in Paper~1 for the oblique shock (solid red line) with the empirical formula, and those estimated here (triangles) with a completely different technique.
} \label{fig05}
\end{figure}

\begin{figure}[th]
\epsscale{0.5}
\plotone{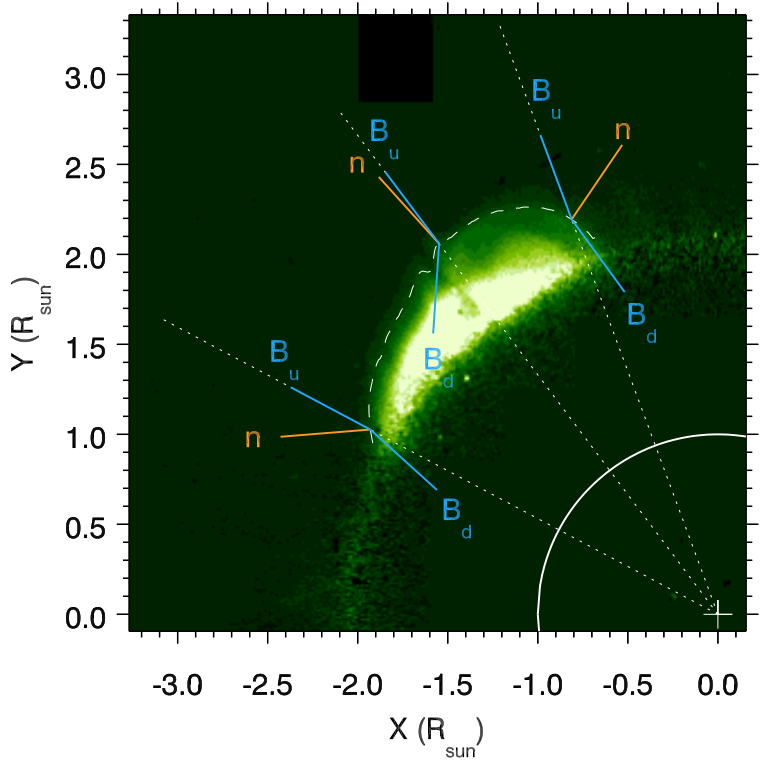}
\newline
\newline
\newline
\caption{
Measured deflections of the up-stream magnetic field (blue solid lines, assumed to be radial before the shock transit) with respect to the shock normal (solid orange lines) for the three points analyzed here along the front (white dashed line).
} \label{fig06}
\end{figure}

\begin{figure}[th]
\epsscale{1.0}
\plotone{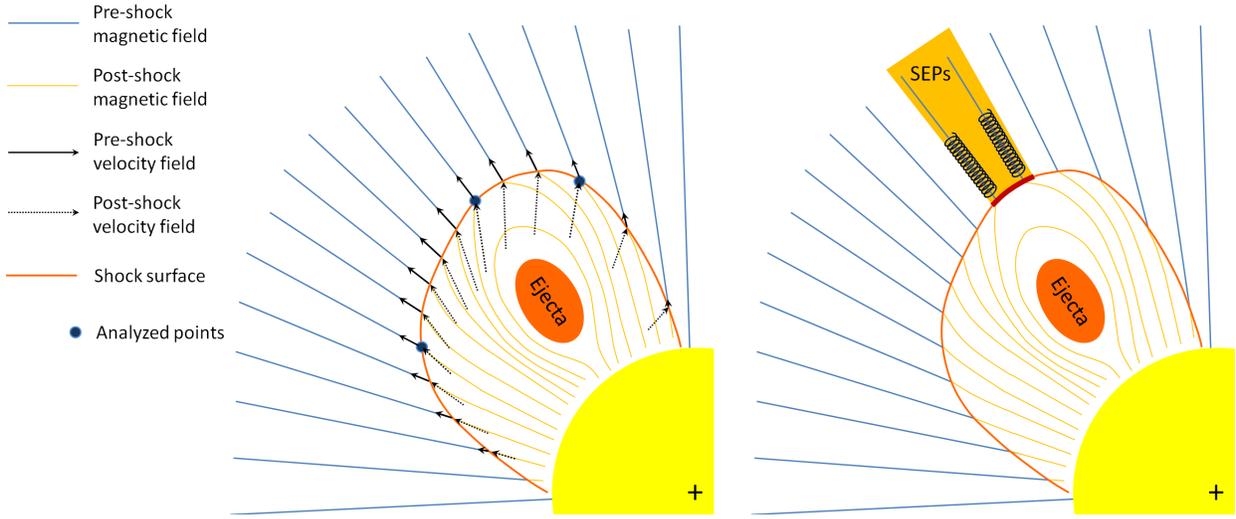}
\caption{
Left: a cartoon showing a possible 2D distribution of magnetic fieldlines on the plane of the sky (POS): the shock surface mimics the one observed in the LASCO~C2 frame (Figure~\ref{fig06}), and pre-shock magnetic and velocity vectors are drawn along radials (as assumed in the computation). Inclinations of post-shock magnetic and velocity vectors with respect to the shock surface corresponds to what was measured for the three points analyzed here (blue filled circles), while inclinations of vectors at other locations along the front have been drawn by imposing continuity. It is evident that the shock transit significantly modifies both the magnetic and velocity fields being met by the expanding ejecta. Right: same as left, showing the latitudinal region (red line) of the shock surface (orange line) where the shock is supercritical (Figure~\ref{fig05}): this likely identifies the coronal region (light orange) where seed particles for SEP acceleration were located at that time \citep[see also][Figure 2, right panel]{ko2013}.  
} \label{fig07}
\end{figure}

In this work we demonstrate that UV and WL data can be combined to derive unique information on the interaction between the coronal plasma and shock waves. This analysis allows us to derive not only the strength of the pre-shock magnetic field at the shock nose \citep[as done by][through the analysis of WL data alone]{gopalswamyyashiro2011}, but also the strength of the pre-shock field at the flanks, the strength of the post-shock field at the nose and at the flanks, together with the rotation of the field vector induced by the transit of the shock itself. In fact, this analysis can be performed not only at the center of the shock, but all along its front, thus allowing the determination of coronal fields at different latitudes and altitudes at the same time. The main results of this work can be summarized as follows:
\begin{itemize}
\item Analysis of WL coronagraphic images (SOHO/LASCO) can be employed to derive not only the up-stream coronal plasma density and the shock compression ratio $X$, but also the angle $\theta_{sh}$ between the normal to the shock front and the radial direction, and the shock speed $v_{sh}$, together with an approximate estimate of the shock Mach number $M_{A\angle}$ for the general case of an oblique shock. All these parameters have been derived here (and in previous Paper~1 and Paper~2) all along the shock front, and hence at different latitudes and altitudes in the corona.
\item Resulting shock speed $v_{sh}$ is larger as expected at the center of the shock ($v_{sh} \sim 1570 - 1580$~km~s$^{-1}$), and then it decreases towards the shock flanks ($v_{sh} \sim 1340 - 1350$~km~s$^{-1}$ about $15^\circ - 25^\circ$ away from the center). Shock compression ratio $X$ and Mach number $M_A$ also are maxima at the shock nose (where $X \simeq 3.0$, $M_{A\angle} \simeq 1.8$) and decrease towards the shock flanks (where $X \simeq 1.2$, $M_{A\angle} \simeq 1.1$ about $15^\circ - 25^\circ$ away from the center).
\item Analysis of UV data (SOHO/UVCS) can be employed to derive the plasma physical parameters missing from the analysis of the WL data: the pre-shock plasma temperature $T$ and outflow velocity $v_{out}$. This work focused on the three coronal points where UV and WL data were available at the same locations in the pre-shock corona: two points at the northward and southward shock flanks and one point at the shock nose.
\item Resulting pre-shock temperatures and velocities are around $T \sim 1.0-1.7 \cdot 10^6$~K and $v_{out} \sim 40-70$~km~s$^{-1}$, consistent with values expected in the analyzed range of heliocentric distances ($2.2-2.6$~R$_\odot$) and latitudes ($30^\circ - 70^\circ$ N). These parameters have been derived from the observed UV (\ovi\ $\lambda$ 1031.91 \AA\ and \hi\ \lya\ $\lambda$ 1215.67 \AA) integrated line intensities alone. Hence, no spectroscopic information are required (e.g. line FWHM and line centroid) in order to repeat this analysis.
\item The above results from the WL and UV data can be combined in order to derive (with MHD RH equations) the full set of post-shock plasma parameters, including the pre- ($B_u$) and post-shock ($B_d$) magnetic field strengths, post-shock outflow velocity, together with the magnetic and velocity field vector rotation angles across the shock surface.
\item Resulting pre-shock coronal magnetic field is around $B_u \simeq 0.2-0.5$~G, hence compatible with values expected in the analyzed range of heliocentric distances ($2.2-2.6$~R$_\odot$), with a latitudinal variation by a factor $\simeq 2$ between the northward and southward coronal point. The Alfv\'en speed $v_A$ is around $v_A \simeq 810-820$~km~s$^{-1}$ at 2.2~R$_\odot$ and increases up to $v_A \simeq 1090$~km~s$^{-1}$ at 2.6~R$_\odot$, with a plasma $\beta \sim 0.01-0.04$.
\item The shock transit corresponds to a magnetic field compression by a factor $\simeq 1.6-1.7$ at the northward flank and the nose, while a weaker compression by a factor $\simeq 1.2$ occurs at the southward flank. Nevertheless, the stronger field rotation occurs at the shock nose, where the field is deflected by $\simeq 46^\circ$ and stronger plasma compression (factor $\sim 2.7$) and heating (factor $\sim 12$) occur. Weaker deflections by $\simeq 14-16^\circ$ occur at the flanks, where more moderate, but still significant (factor $1.5-3.0$) plasma heatings occur. Magnetic field deflections along the shock front are plotted in Figure~\ref{fig06}.
\item Shock Mach numbers $M_A$ measured from the combined WL and UV data analysis are in good agreement with $M_{A\angle}$ values estimated from WL data alone with an empirical formula (Equation~1), which is then validated here. Shock Mach numbers at the nose (flanks) are very close to those expected for a parallel (perpendicular) shock. Hence, shock conditions at the nose (flanks) are very well approximated by a parallel (perpendicular) shock. We also confirm that the shock is super-critical ($M_A \sim M_{A\angle} > M_A^*$) at the nose and sub-critical ($M_A \sim M_{A\angle} < M_A^*$) at the flanks.
\item The shock transit induces a clockwise rotation of the magnetic field vector at the southward flank and at the nose, while a counter-clockwise rotation occurs at the northward flank. This results in a draping of magnetic field lines around the expanding CME. On the other hand, the clockwise rotation of the velocity field vector occurs at both the flanks and also at the nose, resulting in an asymmetric post-shock velocity field being met by the expanding CME (Figure~\ref{fig06}).
\end{itemize}
In order to better describe the above results, we have drawn a cartoon (Figure~\ref{fig07}, left panel) showing the overall possible distribution of pre- and post-shock magnetic and velocity fields all along the shock front. The cartoon is drawn starting from the vector rotations derived for the three points considered in this analysis (Figure~\ref{fig06} and blue filled circles in Figure~\ref{fig07}), and then by assuming continuity all over other latitudes. The resulting magnetic field deflections due to the shock transit correspond to a draping of field lines around the expanding flux rope. Very interestingly, this result is in good agreement with post-shock magnetic field rotations recently obtained by \citet[][]{liu2011} with a 3D MHD numerical simulation. In addition, these authors found that in the CME sheat regions closer to the shock surface (what they call layer~1) ``the magnetic field lines remain in the coplanarity layer as if they are unaffected by the draping field line''. This means that if the pre-shock field lines were lying mainly on POS, the same would also be true for post-shock field lines, hence strongly supporting our assumption that measured magnetic field deflections occur mainly on that plane. It is also interesting to note that the asymmetry in the deflection of velocity vectors (Figure~\ref{fig07}, left panel) is in agreement with the asymmetry of the shock front shape which is also expanding northward in latitude (Figure~\ref{fig02}, left panel). This cartoon evidences, in general, how the physical parameters of post-shock plasma strongly depend not only on the pre-shock magnetic and velocity fields inclination with respect to the shock surface, but also on the shock compression ratio. Moreover, Figure~\ref{fig07} (right panel) shows where the shock is sub- and super-critical, hence the expected coronal regions where seed particles for SEP acceleration could be located. Because, as determined in Paper~1 and Paper~2, the fraction of the front where the shock is super-critical decreases as the shock expands in the corona (see Figure 3 in Paper~1), broader or narrower coronal regions could serve as SEP sources at different times.

As anticipated in the description of the data analysis, plasma temperatures $T$ derived from UV line intensities are representative of electron temperatures $T_e$, mainly because collisions with electrons are responsible for both the determination of the atomic ionization stage and for the collisional excitation of coronal ions. Electron temperatures have been used here (and in previous Paper~3) as input plasma temperatures for the MHD RH equations, thus assuming that electrons and protons are both heated across the shock discontinuity. Nevertheless, as recently pointed out by \citet[][]{manchester2012}, the thermodynamics for the protons and electrons are expected to be different because ``the shock is only supersonic relative to the proton fast-mode speed and not that of the electrons'', hence ``protons receive the kinetic energy dissipated at the shock, while electrons are only heated by their adiabatic compression at the shock''. This is expected to be true also in the special case of the event reported here, because the thermal speed of electrons ($v_e \sim 6740$~km~s$^{-1}$ for $T_e =1.5\cdot 10^6$~K) is much larger than the measured shock speed $v_{sh}$, while the proton thermal speed ($v_p \sim 160$~km~s$^{-1}$ for $T_p =1.5\cdot 10^6$~K) is much smaller than $v_{sh}$ (see Figure~\ref{fig02}, right panel), hence we expect only protons to be directly heated by the shock.

This means that, even if the pre-shock corona is close to thermodynamic equilibrium with $T_p \sim T_e \sim 1.5 \cdot 10^6$ K, the transit of the shock will cause a decoupling between electron and proton temperatures, with $T_p > T_e$ after the transit of the shock. For this reason, in Table~\ref{tab03} we also provided the expected values for down-stream plasma temperatures $T_{d\gamma}$ one could expect from simple adiabatic compression of the considered particles. According to \citet[][]{manchester2012}, we suggest that the down-stream plasma temperatures derived here with RH equations ($T_d$ in Table~\ref{tab03}) could be more representative of post-shock proton temperatures ($T_p \simeq T_d$), while temperatures given by adiabatic compression ($T_{d\gamma}$ in Table~\ref{tab03}) could be more representative of post-shock electron temperatures ($T_{d\gamma} \sim T_e$). In this interpretation, not only our proton temperature increases (by a factor $\sim 12$ at the shock nose and by a factor $1.5-3.0$ at the flanks), but our electron temperature also increases (by a factor $\sim 2$ at the shock nose and by a factor $1.2-1.5$ at the flanks) across the shock are in good agreement with simulation results by \citet[][]{manchester2012} above 1.5~R$_\odot$. After the transit of the shock, electrons and protons are only weakly coupled by collisions: their energy equipartition time $\tau_{pe}$ \citep[as we estimated with formula given by][]{spitzer1962} is on the order of $\tau_{pe} \sim$ 18~hours at the shock nose, hence the post-shock plasma located in the CME sheat and being met by the expanding flux rope will have protons and electrons with significantly different temperatures. To our knowledge, this is the first time that such a deep knowledge of plasma physical parameters across an interplanetary shock is provided from remote sensing data.

This work (and previous analysis in Paper~3) also demonstrates that UV and WL data relative to interplanetary shocks can be combined in order to derive reliable measurements of coronal magnetic fields. In particular, field values derived here are in very good agreement with those provided in the same range of heliocentric distances ($2.6-2.2$~R$_\odot$) with empirical models by \citet[][]{dulkmclean1978} ($\sim 0.25-0.38$~G), \citet[][]{patzold1987} ($\sim 0.51-0.81$~G) and more recently by \citet[][]{mancusogarzelli2013} ($\sim 0.48-0.75$~G). As recently shown by \citet[][]{gopalswamyyashiro2011}, coronal fields can be inferred simply from the analysis of WL coronagraphic images of CME-driven shocks, but when this technique is applied to intensities observed above heliocentric distances of $\sim 2$~R$_\odot$, it requires the assumption of the unknown solar wind velocity, otherwise negligible at lower altitudes \citep[e.g.][]{gopalswamy2012}. On the other hand, we showed here how pre-shock plasma parameters were derived from UV data (electron temperature, density and outflow velocity), and other shock parameters were derived independently from WL data (shock compression ratio, shock velocity and inclination of shock front surface); no coronal physical parameters were assumed. 

As mentioned in the Introduction, before this work, information on shock heating of heavy ions was derived from the analysis of UVCS data by measuring the broadening of spectral lines whose emission is also due to collisional excitation (in particular the \ovi $\lambda$ 1031.91 \AA\ line), because significant dimming of the radiative components is expected after the shock transit. A direct measurement of the post-shock proton temperatures is not straightforward: neutral H atoms do not directly feel the transit of the shock wave, but only indirectly through collisions with post-shock accelerated and heated electrons and proton populations. This will significantly increase the ionization rates due to collisions with electrons and to resonant charge transfer with accelerated protons, thus significantly reducing the fraction of neutrals and producing H atoms traveling with the velocity of post-shock plasma, whose \lya\ emission (due to radiative excitation alone) is thus subject to severe Doppler dimming. Hence the detection of this post-shock faint emission is possible \citep[e.g.][]{mancuso2002}, but not simple. Moreover (as we pointed out) the reliability of $T_p$ measurements from \hi\ \lya\ profiles was also questioned \citep[][]{labrosse2006}. For this analysis only the pre-shock integrated intensities of UV lines were employed, hence no additional information that could be provided by UV spectroscopic data (e.g. spectral line broadening, line Doppler shifts, etc...) was required. Hence, this technique seems very promising for future application to UV (\hi\ \lya) and WL coronagraphic images that will be provided by the METIS coronagraph \citep[][]{antonucci2012} onboard the ESA--Solar Orbiter mission, due to launch in 2017-2018. The technique described here could also be combined with that proposed by \citet[][]{gopalswamyyashiro2011} in order to provide (by imposing the same magnetic field values) an interesting new method to estimate the solar wind velocities at different heliocentric distances in the corona being crossed by CME-driven shocks.

The observational evidence reported here will be used in a validation effort of simulation results in a future paper. The observed shock is closer to the Sun than the most commonly used codes for space weather forecast \citep[e.g. ENLIL initiates its computational domain at 21.5 or 30 R$_\odot$,][]{xie2004}. The present results instead allow one to test the validity of models of CME eruption and production of shocks at closer distances to the Sun. The needed model must be inward of the sonic point in a stratified atmosphere where the density, magnetic field and pressure are chosen according to observational determinations of the average properties of the solar corona. The CME can then be initiated and the ensuing shock can be modeled. The results of the model can then be compared with observational evidence reported here, focusing especially on the dependence of the shock speed on the relative angle with the magnetic field (assumed initially radial).

\acknowledgments
The research leading to these results has received funding from the European Commission Seventh Framework Programme (FP7/2007-2013) under the grant agreement SWIFF (project no. 263340, www.swiff.eu). A.~B. thanks S.~Mancuso for useful discussions in the first phases of this work.

\clearpage

\clearpage

\clearpage

\clearpage

\clearpage

\clearpage

\clearpage

\end{document}